\documentclass[12pt]{article}\usepackage{lady_vanishes}

\begin{document}

\title{Qubiter Algorithm Modification,
Expressing Unstructured Unitary Matrices\\
with Fewer CNOTs}

\author{Robert R. Tucci\\
        P.O. Box 226\\
        Bedford,  MA   01730\\
        tucci@ar-tiste.com}

\date{ \today}

\maketitle

\vskip2cm
\section*{Abstract}
A quantum compiler is  a
software program for
decomposing (``compiling") an
arbitrary unitary matrix into
a sequence of elementary operations (SEO).
The author of this paper is also the author of a
quantum compiler called Qubiter.
Qubiter  uses
a matrix decomposition called
the Cosine-Sine Decomposition (CSD)
that is well known
in the field of Computational Linear Algebra.
One way of measuring
the efficiency of a quantum compiler is
to measure
the number of CNOTs
it uses to express an unstructured
unitary matrix
(a unitary matrix with no special symmetries).
We will henceforth refer to
this number as $\epsilon$.
In this paper, we show how to
improve $\epsilon$ for
Qubiter so that it matches
the current world record for
$\epsilon$,
which is held by another
quantum compiling algorithm based on
CSD.

\newpage
\section{Introduction}

In quantum computing,
elementary operations are operations that
act on only a few (usually
one or two) qubits. For example, CNOTs and
one-qubit rotations are elementary operations.
A quantum compiling algorithm
is an algorithm for
decomposing (``compiling") an
arbitrary unitary matrix into
a sequence of elementary operations (SEO).
A quantum compiler is a software program
that implements a quantum compiling algorithm.

Henceforth, we will refer to
Ref.\cite{Tuc99} as Tuc99.
Tuc99 gives a quantum compiling algorithm,
implemented in a software program called
Qubiter.
The Tuc99 algorithm uses
a matrix decomposition called
the Cosine-Sine Decomposition (CSD)
that is well known
in the field of Computational Linear Algebra.
Tuc99 uses CSD
in a recursive manner. It
decomposes any unitary
matrix into a sequence
of diagonal unitary matrices
and something called
uniformly controlled U(2) gates.
Tuc99 then expresses
these diagonal unitary matrices
and uniformly controlled U(2) gates
as SEOs of short length.

More recently,
two other groups have
proposed quantum compiling
algorithms based on CSD.
One group, based at the Univ.
of Michigan and NIST,
has published Ref.\cite{Mich04},
henceforth referred to as Mich04.
Another group based at Helsinki
Univ. of Tech.(HUT), has published
Refs.\cite{Hels04a}.
and \cite{Hels04b}, henceforth referred
to as HUT04a and HUT04b, respectively.

One way of measuring
the efficiency of a quantum compiler is
to measure
the number of CNOTs
it uses to express an unstructured
unitary matrix
(a unitary matrix with no special symmetries).
We will henceforth refer to
this number as $\epsilon$.
Although good
quantum compilers
will also require optimizations that
deal with structured matrices,
unstructured matrices are certainly an
important case worthy of attention.
Minimizing the number of CNOTs
is a reasonable goal, since a
CNOT operation (or any 2-qubit
interaction used as a CNOT
surrogate) is
expected to take more time to perform
and to introduce more environmental
noise into the quantum computer
than a one-qubit rotation.
Ref.\cite{bound}
proved
that for unitary matrices
of dimension $2^\nb$ ($\nb=$ number of bits),
$\epsilon\geq \frac{1}{4}(4^\nb-3\nb-1)$.
This lower bound
is achieved for $\nb=2$
by the 3 CNOT circuits first
proposed in Ref.\cite{Vidal}.
It is not known whether
this bound can always be achieved
for $\nb>3$.

The Mich04 and HUT04b algorithms
try to
minimize $\epsilon$.
In this paper, we
propose a modification of
the Tuc99 algorithm which will
henceforth be referred to as Tuc04.
Tuc04 comes in two flavors,
Tuc04(NR)
without
relaxation process,
and Tuc04(R) with relaxation process.
 As the next
table shows, the most efficient
algorithm known at present is
Mich04. HUT04b performs worse
than Mich04. Tuc04(R)
and Mich04 are equally efficient.

\begin{tabular}{|l|l|}
\hline
algorithm & $\epsilon$\\
\hline\hline
Tuc99 & $O(2^{2\nb})$\\
Mich04 & $2^{2\nb-1}-3(2^{\nb-1}) +1=(2^\nb-1)(2^{\nb-1}-1)$\\
HUT04b & $2^{2\nb-1}-(2^{\nb-1})-2= (2^\nb-1)(2^{\nb-1}-1) + 2^\nb +O(1)$\\
Tuc04(NR) & $(2^\nb-1)(2^{\nb-1}-1) + 2^\nb$\\
Tuc04(R) & $(2^\nb-1)(2^{\nb-1}-1)$\\
\hline
\end{tabular}

Caveat: Strictly
speaking, the efficiency of
Tuc04(R)
as listed in this table
is only a conjecture.
The problem is that
Tuc04(R) uses a relaxation process.
This paper
argues, based on intuition, that
the relaxation process converges, but
it does not prove this rigorously.
A rigorous proof of
the efficiency of Tuc04(R) will require
theoretical and numerical
proof that its
relaxation process converges
as expected.

\section{Notation}

This paper is based heavily
on Tuc99 and
assumes that the reader
is familiar with
the main ideas of Tuc99.
Furthermore,
this paper uses the notational
conventions of
Tuc99. So if the reader can't follow
the notation of this paper, he/she
is advised to consult Tuc99.
The section on notation in Ref.
\cite{Paulinesia} is also recommended.

Contrary to Tuc99, in this paper we
 will normalize Hadamard matrices so that
their square equals one.

As in Tuc99, for  a single qubit
with number operator $n$,
we define $P_0 = n$ and
$P_1 = \overline{n} = 1-n$.
If $\vec{\kappa}=(\kappa_1, \kappa_2,
\ldots, \kappa_\nk)$  labels $\nk$
distinct qubits and
$\vec{b}=(b_1, b_2, \ldots, b_{\nk})\in Bool^\nk$,
then we define
$P_{\vecb}(\vec{\kappa}) =
P_{b_1}(\kappa_1)P_{b_2}(\kappa_2)
\ldots P_{b_\nk}(\kappa_\nk)$.

When we say $A$ (ditto, $A'$)
is $B$ (ditto, $B'$), we mean
$A$ is $B$ and $A'$ is $B'$.

For any complex number $z$,
we will write $z= |z| e^{i\angle(z)}$.
Thus, $|z|$ and $\angle(z)$ are
the magnitude and phase angle of $z$,
respectively.

$\hate_x, \hate_y, \hate_z$ will denote
the unit vectors along the X, Y, Z axes,
respectively.
For any 3d real unit vector $\hats$,
$\sigma_s = \vec{\sigma}\cdot \hats$,
where $\vec{\sigma}=(\sigx, \sigy, \sigz)$
is the vector of Pauli matrices.

\section{$U(N)$-subsets and $U(N)$-Multiplexors}

We define a {\bf $U(N)$-subset}
to be an ordered set
$\{U_b\}_{\forall b}$
of $N$ dimensional unitary matrices.
Let the index
$b$ take values in a set $S_\rvb$ with $N_\rvb$
elements.
In this
paper, we are
mostly concerned with
the case that $S_\rvb = Bool^{\nk}$,
and $b$ is represented by $\vec{b}$.

\begin{figure}[h]
    \begin{center}
    \epsfig{file=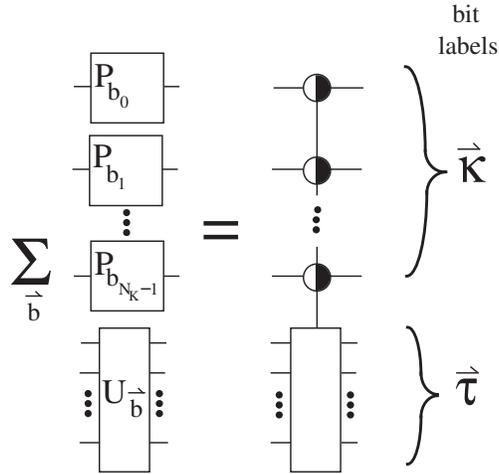, height=2.5in}
    \caption{2 Diagrammatic representations
    of a $U(2^\nt)$-multiplexor.}
    \label{fig-multiplexor}
    \end{center}
\end{figure}

Suppose
a qubit array with $\nb$
qubits is partitioned
into $\nt$ target
qubits and $\nk$ control qubits.
Thus, $\nt, \nk$ are
positive integers such that $\nb=\nt+\nk$.
Let $\vec{\kappa}=(\kappa_1, \kappa_2,
\ldots, \kappa_\nk)$ denote the control
qubits and
$\vec{\tau}=(\tau_1, \tau_2,
\ldots, \tau_\nt)$
the target qubits.
Thus, if $\vec{\tau}$ and $\vec{\kappa}$
are considered as sets,
they are disjoint and their union is
$\{0, 1, \ldots, \nb-1\}$.
Let
$\{U_\vecb\}_{\forall \vecb\in Bool^\nk}$
be an ordered set of operators all of which
act on the Hilbert space of the target qubits.
We will refer to any operator
$X$ of the following
form as a
{\bf uniformly controlled
$U(2^\nt)$-subset}, or, more succinctly,
as
a {\bf $U(2^\nt)$-multiplexor}:

\beq
X =
\sum_{\vecb\in Bool^\nk}
P_{\vecb}(\vec{\kappa}) U_\vecb(\vec{\tau})
=
\prod_{\vecb\in Bool^\nk}
U_\vecb(\vec{\tau})^{P_{\vecb}(\vec{\kappa})}
\;.
\eeq
(``multiplexor" means ``multi-fold" in Latin.
A special type of electronic device
is commonly called a multiplexor or multiplexer).
Note that
$X$ is a function of:
a set $\vec{\kappa}$
of control bits,
a set $\vec{\tau}$
of target bits, and
a $U(2^\nt)$-subset
$\{U_\vecb\}_{\forall \vecb\in Bool^\nk}$.
Fig.\ref{fig-multiplexor} shows two possible
diagrammatic representations
of a multiplexor, one more
explicit than the other.
The
diagrammatic representation
with the ``half moon" nodes
was introduced in Ref.\cite{Hels04a}.

For a given
$U(2)$-subset $\{U_b\}_{\forall b}$ (and for
any multiplexor with
that $U(2)$-subset),
it is useful to define
as follows
what we shall call
the optimal axis of the
$U(2)$-subset.
Suppose that we express each
$U_b$ in the form

\beq
U_b = e^{i\eta_b}
e^{i\sigz \gamma_b}
e^{i(\sigsone \alpha_b
+
\sigstwo \beta_b)}
(i\sigw)^{f(b)}
\;,
\label{eq-parametri-left-diag}
\eeq
where $\eta_b, \alpha_b, \beta_b, \gamma_b$
are real parameters, where the vectors
$\hats_1, \hats_2$, and
$\hatw=\hats_1\times\hats_2$ are orthonormal,
and where $f(b)$ is an indicator
function which maps the
set of all possible $b$ into $\{0,1\}$.
Of course, $e^{i\eta_b} = \sqrt{\det{U_b}}$.
Appendix
\ref{app-param} shows how to
find the parameters
$\alpha_b, \beta_b, \gamma_b$ for a given
$U_b e^{-i\eta_b}\in SU(2)$.
Appendix \ref{app-mini}
solves the following minimization problem.
If the value of the parameters
$\alpha_b, \beta_b, \gamma_b$
and the vectors $\hats_1, \hats_2, \hatw$
are allowed to vary, while keeping
the vectors $\hats_1, \hats_2, \hatw$ orthonormal
and keeping all $U_b$ fixed,
find vectors $\hats_1, \hats_2, \hatw$
that are optimal, in the sense
that they minimize
a cost function.
The cost function
penalizes deviations of the diagonal
matrices $e^{i\gamma_b\sigz}$
away from the 2d identity matrix
$I_2$.
Any choice of
orthonormal vectors
$\hats_1, \hats_2$
will be called {\bf strong directions}
and $\hatw=\hats_1\times\hats_2$
will be called a {\bf weak direction},
or an
{\bf axis of the $U(2)$-subset}.
An axis that minimizes the cost function
will be called
the {\bf optimum axis of the $U(2)$-subset}.
(an axis of
goodness).

It is also possible to define
an optimum axis of a $U(2)$-subset
in the same way as just discussed,
except replacing
Eq.(\ref{eq-parametri-left-diag})
by

\beq
U_b = e^{i\eta_b}
(i\sigw)^{f(b)}
e^{i(\sigsone \alpha_b
+
\sigstwo \beta_b)}e^{i\sigz \gamma_b}
\;.
\label{eq-parametri-right-diag}
\eeq
In
Eq.(\ref{eq-parametri-left-diag}),
the diagonal matrix
$e^{i \gamma \sigz}$ is on the left
hand side, so we will call this
the {\bf diagonal-on-left (DOL)
parameterization}.
In
Eq.(\ref{eq-parametri-right-diag}),
the diagonal matrix
$e^{i \gamma \sigz}$ is on the right
hand side, and we will call this
the {\bf diagonal-on-right (DOR)
parameterization}.

\section{Tuc04 algorithm}
\begin{figure}[h]
    \begin{center}
    \epsfig{file=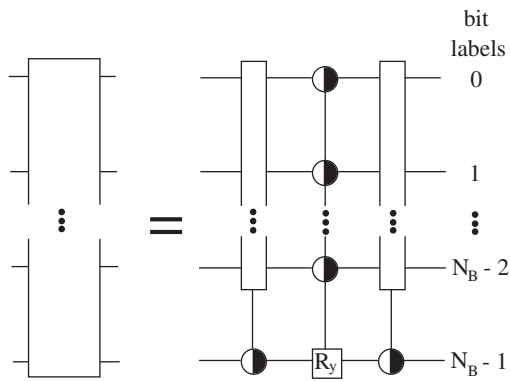, height=2.0in}
    \caption{Diagrammatic representation
    of the Cosine-Sine Decomposition (CSD).}
    \label{fig-CSD}
    \end{center}
\end{figure}

\begin{figure}[h]
    \begin{center}
    \epsfig{file=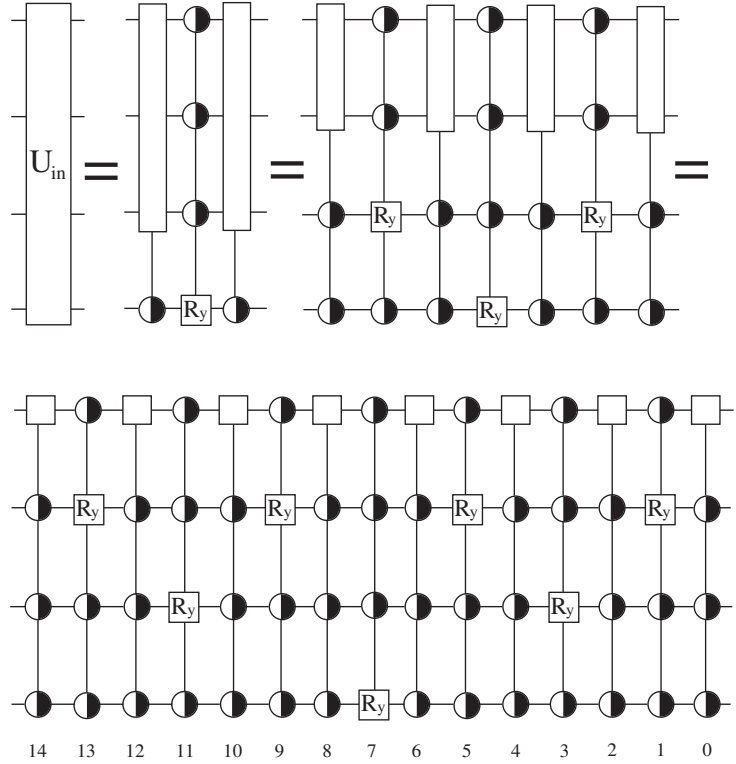, height=4.0in}
    \caption{Recursive use of CSD to
    decompose a $2^4$ dimensional unitary matrix.}
    \label{fig-Qubiter-4bits}
    \end{center}
\end{figure}

The Cosine Sine Decomposition
(CSD) expresses
an $N$ dimensional unitary
matrix $U$ as a product
$LDR$, where
$L=L_0\oplus L_1$,
$D=e^{i\sigy\otimes \Theta}$,
$R=R_0\oplus R_1$,
where
$ L_0, L_1, R_0, R_1$
are unitary matrices of
dimension $N/2$,
and $\Theta$
is a diagonal real matrix
whose entries can be interpreted as
angles between subspaces.
Note that
the matrices $L, D$ and $R$
are all multiplexors.
Fig.\ref{fig-CSD} depicts the CSD
graphically, using the
multiplexor symbol of Fig.\ref{fig-multiplexor}.
In Fig.\ref{fig-CSD},
a $U(2)$-multiplexor
whose $U(2)$-subset
consists
solely of
rotations around
the Y axis, is indicated
by putting the symbol $R_y$
in its target box.
We will call this type
of multiplexor an {\bf $R_y(2)$-multiplexor}.

Lets review the Tuc99 algorithm.
It decomposes an
arbitrary unitary matrix
into a SEO
by applying the
CSD in a recursive manner.
The beginning of the
Tuc99 algorithm for $\nb=4$
is illustrated
in Fig.\ref{fig-Qubiter-4bits}.
An initial
unitary matrix $U_{in}$
is decomposed via CSD into a product
of 3 multiplexors $L,D,R$. The
$L$ and $R$ multiplexors
on each side of $D$
are in turn decomposed via CSD.
The $L$ and $R$
multiplexors generated via any
application of CSD are in turn
decomposed via CSD.
In Fig.\ref{fig-Qubiter-4bits}, we have
stopped recursing once
we reached multiplexors
whose target box acts on a single
qubit. Note
that at this stage,
$U_{in}$ is decomposed
into a product of $U(2)$-multiplexors.
There are $1+2+4+\ldots+2^{\nb-1}=2^{\nb}-1$
of these $U(2)$-multiplexors
(15 for $\nb=4$). Half of these
$U(2)$-multiplexors
have $R_y$
in their target boxes and the other
half don't. Furthermore the $R_y$
type multiplexors and non-$R_y$
ones alternate.
Furthermore, the non-$R_y$
$U(2)$-multiplexors have
their target box at qubit 0,
so, according to the conventions
of Tuc99, they are direct sums of
$U(2)$ matrices.
The
Tuc99 algorithm
deals with these
direct sums of $U(2)$ matrices
by applying CSD to each $U(2)$ matrix
in the direct sum.
This converts each direct
sum of $U(2)$ matrices into
a product $LDR$, where
$R$ and $L$ are diagonal unitary matrices
and $D$ is an $R_y(2)$-multiplexor.
Thus,  Tuc99
turns the last operator
sequence shown in Fig.\ref{fig-Qubiter-4bits}
into a sequence of alternating
diagonal unitary matrices and
$R_y(2)$-multiplexors.
Then Tuc99
gives a prescription for
decomposing any diagonal unitary
matrix into a SEO with
$2^{\nb}$ CNOTs
and
any $R_y(2)$-multiplexor
into a SEO with $2^{\nb-1}$ CNOTs.

Tuc99 considers
what it calls a $D$-matrix:

\begin{subequations}
\label{eq-gen-d-def}
\beq
D = \exp(\sum_{\vecb \in Bool^{\nb-1}}
i \Phi_\vecb \otimes P_\vecb)
=
\sum_{\vecb \in Bool^{\nb-1}}
U_\vecb
\otimes P_\vecb
\;,
\eeq
with

\beq
U_\vecb = \exp(i \Phi_\vecb)
\;,
\eeq
\end{subequations}
where

\beq
\Phi_\vecb
=
\phi_\vecb \sigy
\;.
\eeq
Here $\phi_\vecb$ is a real parameter.
In the nomenclature of
this paper, $D$ is an $R_y(2)$-multiplexor
with a single target qubit  at $\nb-1$
and $\nb-1$ control qubits
at
$\{0,1, \ldots,\nb-2\}$.
Tuc99 shows how to decompose
$D$ into a SEO with $2^{\nb-1}$ CNOTs.
Tuc99 also discusses
how, by permuting qubits
via the qubit exchange operator,
one can move the target
qubit to any position
$\{0, 1, \ldots, \nb-1\}$
to get what Tuc99 calls a direct
sum of $D$ matrices.
In the nomenclature of this paper,
a ``direct
sum of $D$ matrices" is just an
$R_y(2)$-multiplexor
with a single target qubit
at any position
out of $\{0, 1, \ldots, \nb-1\}$.
In conclusion, Tuc99
gives a complete
discussion of
$R_y(2)$-multiplexors
and how to decompose them
into a SEO with $2^{\nb-1}$ CNOTs.

Next, let us consider how
to generalize Tuc99.
We begin by
proving certain
facts about
$U(2)$-multiplexors
that are generalizations of
similar facts obtained in
Tuc99
for
$R_y(2)$-multiplexors.

Suppose $\hats_1, \hats_2$
and
$\hatw=\hats_1\times \hats_2$ are
orthonormal vectors.
Suppose we generalize the
$D$ matrices of Tuc99
by using Eqs.(\ref{eq-gen-d-def})
with:

\beq
\Phi_\vecb
=
\phi_{\vecb,1} \sigsone
+
\phi_{\vecb,2} \sigstwo
\;.
\label{eq-phi-def-sw}
\eeq
Here
$\phi_{\vecb,x} $ and $\phi_{\vecb,y}$
are real parameters.
In Tuc99, we define
$\vec{\phi}$ to be a column vector
whose components are the numbers
$\phi_\vecb$ lined up in order of
increasing $\vecb\in Bool^{\nb-1}$.
Here, we use the same
rule to define vectors $\vec{\phi}_x$
and $\vec{\phi}_y$
from $\phi_{\vecb,x} $ and $\phi_{\vecb,y}$,
respectively. In analogy
with Tuc99, we then define
$\vec{\theta}_x$
and $\vec{\theta}_y$ via
a Hadamard Transform:

\beq
\vec{\theta}_j
=
\frac{1}
{\sqrt{2^{\nb-1}}}
\;\;
H_{\nb-1}
\vec{\phi}_j
\;
\eeq
for
$j\in \{ x, y\}$. ($H_{\nb-1}$ has been
normalized so its square equals one).
Let

\beq
\Theta_\vecb=
\theta_{\vecb,1} \sigsone+
\theta_{\vecb,2} \sigstwo
\;.
\eeq
As in Tuc99,
$D$ can be expressed as

\beq
D = \prod_{\vecb\in Bool^{\nb-1}}
A_\vecb
\;,
\eeq
where the
operators $\{A_\vecb\}_{\forall \vecb}$
mutually commute, and
can be expressed as

\beq
A_\vecb=
\exp\left(i \Theta_\vecb(\nb-1)
\prod_{j=0}^{r-1}\sigz(\beta_j)
\right)
\;.
\label{eq-a-in-exp-form}
\eeq
Next we will use the
following CNOT identities.
For any two distinct bits $\alpha, \beta$,

\beq
\sigw(\bita)^{n(\bitb)}\odot \sigsone(\bita) =
\sigsone(\bita) \sigz(\bitb)
\;,
\eeq
and

\beq
\sigw(\bita)^{n(\bitb)}\odot \sigstwo(\bita) =
\sigstwo(\bita) \sigz(\bitb)
\;.
\eeq
These CNOT identities are easily proven
by checking them separately for the two
cases
$n(\bitb)=0$ and $n(\bitb)=1$.
By virtue of these CNOT identities,
Eq.(\ref{eq-a-in-exp-form})
can be re-written as

\beq
A_\vecb=
[
\sigw(\nb-1)^{n(\bitb_{r-1})}
\ldots
\sigw(\nb-1)^{n(\bitb_{1})}
\sigw(\nb-1)^{n(\bitb_{0})}
]
\odot
\exp[
i\Theta_\vecb(\nb-1)]
\;.
\label{eq-Ab-def-basis}
\eeq

As shown in Tuc99, if we
multiply the $A_\vecb$ matrices
(given by Eq.(\ref{eq-Ab-def-basis}))
in a Gray order in $\vecb$, many
$\sigw(\nb-1)^{n(\bitb)}$ cancel.
We end up expressing $D$ as a SEO
wherein one-qubit rotations
(of bit $\nb-1$) and
$\sigw(\nb-1)^{n(\bitb)}$
type operators alternate,
and there is the same number
($2^{\nb-1}$) of each.
At this point,
the $\sigw(\nb-1)^{n(\bitb)}$
operators may be
converted to CNOTs
using:

\beq
\sigw(\nb-1)^{n(\bitb)}=
e^{i\vec{\sigma}(\nb-1)\cdot\vec{\theta}_{wx}}
\odot
\sigx(\nb-1)^{n(\bitb)}
\;,
\eeq
where
$e^{i\vec{\sigma}(\nb-1)\cdot\vec{\theta}_{wx}}$
is a one-qubit rotation that takes direction
$\hatw$ to direction $\hate_x$.

Even for the generalized $D$
discussed here (i.e., for the $D$
with $\Phi_\vecb$ defined
by Eq.(\ref{eq-phi-def-sw})),
it is still true that,
by permuting qubits
via the qubit exchange operator,
one can move the target
qubit to any position
$\{0, 1, \ldots, \nb-1\}$.

As we have shown,
our generalized $D$ matrix
can be decomposed into
an alternating  product of one-qubit rotations
and CNOTs.
The product contains $2^{\nb-1}$
(one factor of 2 for each control qubit)
CNOTs and the same number of
one-qubit rotations.
This product expression
for $D$ will contain
a CNOT at the beginning
and a one-qubit rotation at
the end, or vice versa, whichever
we choose. Suppose we choose
to have a CNOT at the beginning of
the product, and
that this CNOT is $\sigw(\nb-1)^{n(\mu)}$,
for some $\mu\in\{0,1, \ldots, \nb-2\}$.
Then the matrix
$D[i\sigw(\nb-1)]^{n(\mu)}$
can be expressed with one CNOT less
than $D$,
as a product which starts and
ends with a one-qubit rotation.
And $D[i\sigw(\nb-1)]^{n(\mu)}$
is a $U(2)$-multiplexor just as
much as $D$ is. Indeed,

\begin{subequations}
\beqa
[i\sigw(\nb-1)]^{n(\mu)}&=&
i\sigw(\nb-1)n(\mu) + \overline{n}(\mu)\\
&=&
i\sigw(\nb-1)P_0(\mu) + P_1(\mu)
\;,
\eeqa
\end{subequations}
so

\begin{subequations}
\beqa
D[i\sigw(\nb-1)]^{n(\mu)}&=&
 [\sum_\vecb e^{i\Phi_\vecb}\otimes
P_\vecb][i\sigw(\nb-1)]^{n(\mu)}\\
&=&
\sum_{\vecb\in S_0(\mu)}
(e^{i\Phi_\vecb} i\sigw)\otimes P_\vecb
+
\sum_{\vecb\in S_1(\mu)}
e^{i\Phi_\vecb}\otimes P_\vecb\\
&=&
\sum_\vecb
[e^{i\Phi_\vecb}(i\sigw)^{1-b_\mu}]\otimes P_\vecb
\;,
\eeqa
\end{subequations}
where $S_0(\mu)=\{\vecb\in Bool^{\nb-1}:b_\mu=0\}$
and $S_1(\mu)$ is the complement of $S_0(\mu)$.
Thus, the $U(2)$-subset
of
$D[i\sigw(\nb-1)]^{n(\mu)}$
is the same as that of $D$ except that
half of the $e^{i\Phi_\vecb}$ matrices are
multiplied by $i\sigw$.

In conclusion,
we have
pointed out
a convenient type of
$U(2)$-multiplexor.
The $U(2)$-subset
of a {\bf convenient
$U(2)$-multiplexor}
consists of matrices of the
form
$e^{i\Phi_\vecb}(i\sigw)^{f(\vecb)}$,
where $\Phi_\vecb$ is given
by Eq.(\ref{eq-phi-def-sw})
and $f$ is an indicator function
that maps the set of all $\vecb$ into $Bool$.
A convenient
$U(2)$-multiplexor can be expressed as
a SEO with $2^{\nb-1}-1$
CNOTs.

Next we will give an
algorithm that
converts a $U(2)$-multiplexor
 sequence such as the last operator sequence
in Fig.\ref{fig-Qubiter-4bits}
into a sequence of
convenient $U(2)$-multiplexors.
For definiteness,
we will describe the algorithm
assuming $\nb=4$. How to generalize
the algorithm
to arbitrary $\nb$ will be obvious.

\begin{enumerate}
\item As in Fig.(\ref{fig-Qubiter-4bits}),
let $U_{in}$ be the matrix
to which CSD is initial applied.
We assume that before we start
applying CSD, $U_{in}$
has been normalized so that
$\det(U_{in})=1$.

\item Apply CSD recursively,
as show in Fig.\ref{fig-Qubiter-4bits}.
Let $\Upsilon_j$,
where $0\leq j \leq 14$,
denote the 15 $U(2)$-multiplexors
labelled 0 thru 14
in Fig.\ref{fig-Qubiter-4bits}.
Thus,
$U_{in}= \Upsilon_{14}
\ldots\Upsilon_2\Upsilon_1\Upsilon_0$.

\item \label{step-axis}
For now, let $\{U_b\}_{\forall b}$ denote
the $U(2)$-subset of the
multiplexor $\Upsilon_0$ .
Find
the optimum axis
of $\{U_b\}_{\forall b}$
when the $U_b$ are expressed
in the DOL form:
$U_b = e^{i\eta_b}e^{i\sigz \gamma_b}
e^{i(\sigsone \alpha_b
+
\sigstwo \beta_b)}(i\sigw)^{f(b)}$.
Note that $\Upsilon_0 = \Delta_0
\Upsilon^{conv}_0$, where
$\Upsilon^{conv}_0$ is a
convenient $U(2)$-multiplexor,
and $\Delta_0$ is a diagonal
unitary matrix that incorporates
the diagonal matrix factor
$e^{i\eta_b}e^{i\sigz \gamma_b}$
of each $b$.
Now define the
``intermediate" matrix
$\Upsilon^{inter}_1=\Upsilon_1\Delta_0$.
Note that $\Upsilon^{inter}_1$
is a $U(2)$-multiplexor.
In general, the product of
a $U(2)$-multiplexor times a
diagonal unitary matrix is again
a $U(2)$-multiplexor.

\item For $j=1, 2, \dots 13$, process
$\Upsilon^{inter}_j$ in
the same way that
$\Upsilon_0$ was processed.
In other words, find
the optimum axis
(for a DOL parametrization)
of the $U(2)$-subset of $\Upsilon^{inter}_j$.
Note that
$\Upsilon^{inter}_j = \Delta_j
\Upsilon^{conv}_j$, where
$\Upsilon^{conv}_j$ is a
convenient $U(2)$-multiplexor,
and $\Delta_j$ is a diagonal
unitary matrix.
Now define the
 matrix
$\Upsilon^{inter}_{j+1}=\Upsilon_{j+1}\Delta_j$.
\end{enumerate}

After applying the
previous steps,
we will be able to
write
$U_{in}= \Upsilon_{14}^{inter} \Upsilon_{13}^{conv}
\ldots\Upsilon_1^{conv}\Upsilon_0^{conv}$.
In this expansion of $U_{in}$, all except
the last multiplexor are
of the convenient type.

\begin{enumerate}
\item[$5.$]
One possibility at this point
is to process
$\Upsilon_{14}^{inter}$ and then stop.
That is, express
$\Upsilon_{14}^{inter}$
as a product of a diagonal
unitary matrix
$\Delta_{14}$ and a convenient
multiplexor
$\Upsilon^{conv}_{14}$.
Then express each of the 15
convenient multiplexors
$\Upsilon^{conv}_j$
for $j=0, 1,\ldots 14$ as
a SEO with
$2^{\nb-1}-1$ CNOTs.
Finally, expand
the diagonal unitary matrix $\Delta_{14}$
as a SEO with $2^{\nb}$ CNOTs,
using the technique given
in Tuc99 for doing this.
\end{enumerate}

\begin{enumerate}
\item[$5'.$]
A second possibility is to
repeat the previous steps in
the reverse direction, this time
going from
left to right,
and using DOR parameterizations.
Continue to sweep
back and
forth across the sequence of multiplexors.
We conjecture that after a few
sweeps,
we will start producing
diagonal matrices
$\Delta_j$ that are closer and closer
to unity.
When the latest
$\Delta_j$ matrix is acceptably
close to unity, the process
can be stopped.
At this point,
the axes of the multiplexors
will have reached a kind of
equilibrium, and
we will have
expressed
 $U_{in}$
as a product
of convenient
$U(2)$-multiplexors.
\end{enumerate}

Sweeping only once (ditto, many times)
 is what we called
the Tuc04(NR) algorithm (ditto,
the Tuc04(R) algorithm)
in the Introduction section
of this paper.

For Tuc04(R),
$U_{in}$ is expressed as product of
$2^{\nb}-1$ convenient
$U(2)$-multiplexors,
each of which is expressed as
$2^{\nb-1}-1$ CNOTs,
so
$\epsilon= (2^{\nb}-1)(2^{\nb-1}-1)$.

For Tuc04(NR),
 finding the optimum axis
of each $U(2)$-multiplexor
is unnecessary.
Doing so changes the final diagonal matrix
$\Delta_{14}$,
but does not cause it to vanish.
The lady does not vanish.
Thus, for Tuc04(NR),
it is best to simply use
$(\hats_1,\hats_2, \hatw)
=(\hate_x, \hate_y, \hate_z)$
throughout.
The Tuc04(NR) algorithm
is essentially
the same as the
HUT04b algorithm.
Tuc04(NR),
compared with Tuc04(R),
has the penalty
of having to expand the
final diagonal
matrix $\Delta_{14}$.
This produces an extra
$2^\nb$ CNOTs. So for
Tuc04(NR),
$\epsilon= (2^{\nb}-1)(2^{\nb-1}-1)
+ 2^\nb$.

Note that for Tuc04(R),
it is not necessary to
find very precisely
the optimum axis
of each $U(2)$-multiplexor.
Any errors in finding
such an axis do not
increase the numerical
errors of compiling $U_{in}$.
It may even be true that
the axes equilibrate as long as
one provides, each time step \ref{step-axis}
above calls for an
axis of a U(2)-multiplexor,
an axis that has a better than
random chance of
decreasing the cost function
defined in Appendix \ref{app-mini}.

\appendix
\section{Appendix: Parameterizations of\\
SU(2) matrices}\label{app-param}

In this appendix, we will
show how, given orthonormal
vectors $\hats_1, \hats_2$
and $\hatw=\hats_1\times\hats_2$,
and
given any
SU(2) matrix $U$,
one can find
real parameters $\alpha, \beta, \gamma$
such  that
$U=e^{i \gamma \sigz}
e^{i (\alpha \sigsone  + \beta \sigstwo) }$.
We will
use the well known identity
\beq
e^{i\vec{\sigma}\cdot\vec{\theta} }
=
\cos \theta +
i (\vec{\sigma}\cdot\hat{\theta})
\sin \theta
\;,
\label{eq-funda-id}\eeq
where $\vec{\sigma}=(\sigx, \sigy, \sigz)$,
$\vec{\theta}$ is a real
3d vector of magnitude $\theta$, and
$\hat{\theta} = \vec{\theta}/\theta$.

Note that given a matrix $U\in SU(2)$,
if we express its transpose $U^T$
in the form
$U^T=e^{i \gamma \sigz}
e^{i (\alpha \vec{\sigma}\cdot\hats_1
+ \beta \vec{\sigma}\cdot\hats_2) }$,
then this gives
an expression for
$U$ of the form
$U=
e^{i (\alpha \vec{\sigma}\cdot\hats'_1
+ \beta \vec{\sigma}\cdot\hats'_2) }
e^{i \gamma \sigz}$
where for $j\in\{1,2\}$,
$s_{jx}'=s_{jx}$,
$s_{jy}'=-s_{jy}$,
and
$s_{jz}'=s_{jz}$.
(This follows from the fact
that
$\sigx^T=\sigx$, $\sigy^T=-\sigy$,
$\sigz^T=\sigz$.)
Likewise,
given a matrix $U\in SU(2)$,
if we express $U(-i\sigw)$
in the form
$U(-i\sigw)=e^{i \gamma \sigz}
e^{i (\alpha \sigsone
+ \beta \sigstwo) }$,
then this gives an expression
for $U$ of the form
$U=e^{i \gamma \sigz}
e^{i (\alpha \sigsone
+ \beta \sigstwo) }(i\sigw)$.

In the general case, the
triad $(\hats_1, \hats_2, \hate_z)$
is an oblique (not orthogonal)
basis of real 3d space.
As warm up practice, consider
first the simpler case
when the triad is orthogonal; that is,
when
$\hats_1 = \hate_x$, $\hats_2 = \hate_y$.
Any $U\in SU(2)$ can be expressed as
$\left[
\begin{array}{cc}
x & y \\
-y^* & x^*
\end{array}
\right]$, where
$x, y$ are complex numbers such that
$|x|^2+|y|^2=1$. Thus,
we want to express $\alpha, \beta, \gamma$
in terms of $x,y$, where:
\beq
\left[
\begin{array}{cc}
x & y \\
-y^* & x^*
\end{array}
\right]
=
e^{i \gamma \sigz}
e^{i (\alpha \sigx  + \beta \sigy) }
\;.
\label{eq-left-diag-ort}
\eeq
Let
$\theta = \sqrt{\alpha^2 + \beta^2}$.
Using Eq.(\ref{eq-funda-id}),
it is easy to show that

\begin{subequations}
\label{eq-left-diag-ort-xy}
\beq
x = e^{i \gamma}\cos \theta
\;,
\eeq
and

\beq
y = e^{i\gamma}
\frac{(\beta + i \alpha)}{\theta}
\sin\theta
\;.
\eeq
\end{subequations}
If we assume that
$\cos\theta\geq 0$,
then Eqs.(\ref{eq-left-diag-ort-xy})
can be easily inverted. One finds

\begin{subequations}
\label{eq-left-diag-ort-abc}
\beq
\gamma = \angle(x)
\;,
\eeq

\beq
\cos \theta = |x|
\;,
\eeq
and

\beq
\beta + i \alpha =
\frac{y x^*}{|xy|}\theta
\;.
\eeq
\end{subequations}

Next, we consider the general
case when
the triad $(\hats_1, \hats_2, \hate_z)$
is oblique. One has

\beq
\left[
\begin{array}{cc}
x & y \\
-y^* & x^*
\end{array}
\right]
=
e^{i \gamma \sigz}
e^{i (\alpha \sigsone  + \beta \sigstwo) }
\;.
\eeq
Define $\vec{\theta}$ by

\beq
\vec{\theta} =
\alpha \hats_1 + \beta \hats_2 =
\theta_x \hate_x
+\theta_y \hate_y
+\theta_z \hate_z
\;.
\eeq
Thus,

\beq
\theta = \sqrt{\alpha^2 + \beta^2}
=
\sqrt{\theta_x^2 + \theta_y^2 +\theta_z^2}
\;.
\eeq
Using Eq.(\ref{eq-funda-id}), it is
easy to show that

\begin{subequations}
\label{eq-left-diag-obl}
\beq
x = e^{i\gamma}( \cos\theta
+ i \frac{\theta_z}{\theta} \sin\theta )
\;,
\eeq
and

\beq
y = e^{i \gamma}
\left(
\frac{\theta_y + i \theta_x}{\theta}
\right)
\sin\theta
\;.
\eeq
\end{subequations}
We want to express
$\alpha, \beta, \gamma$
in terms of $x, y$.
Unlike when the triad was
orthogonal,
now expressing $\gamma$ in
terms of $x,y$ is non-trivial;
as we shall see below, it
requires solving numerically for
the root a
non-linear
equation.
The good news is that
if we know $\gamma$,
then $\alpha$ and $\beta$ follow
in a straightforward manner
from:

\beq
\cos \theta = \re (x e^{-i\gamma})
\;,
\eeq
and

\beq
\theta_y+i\theta_x
=(y e^{-i\gamma})\frac{\theta}{\sin \theta}
\;.
\eeq
Given $\theta_x, \theta_y$,
one can find $\alpha, \beta$
using Eq.(\ref{eq-ab-fun-theta}).

Since $|x|^2+|y|^2=1$,
Eqs.(\ref{eq-left-diag-obl})
are equivalent to the
following 3 equations:

\begin{subequations}
\label{eq-abc-theta-xyz-contraints}
\beq
|x|^2 = \cos^2 \theta
+
\left(
\frac{\theta_z}{\theta}
\right)^2
\sin^2\theta
\;,
\eeq

\beq
\angle(x)=
\gamma +
\arctan
\left(
\frac{\theta_z \sin\theta}{\theta \cos\theta}
\right)
\;,
\eeq
and

\beq
\angle(y)=
\gamma +
\arctan
\left(
\frac{\theta_x }{\theta_y}
\right)
\;.
\eeq
As stated previously,

\beq
\vec{\theta} =
\alpha \hats_1 + \beta \hats_2
\;.
\label{eq-def-vec-theta}
\eeq
\end{subequations}
Next, we will
solve the 6 equations given by
Eqs.(\ref{eq-abc-theta-xyz-contraints}) for
the 6 unknowns $(\alpha, \beta, \gamma,
\theta_x, \theta_y, \theta_z)$.

From Eq.(\ref{eq-def-vec-theta}),
it follows that

\beq
\left[
\begin{array}{cc}
s_{1x} & s_{2x} \\
s_{1y} & s_{2y}
\end{array}
\right]
\left[
\begin{array}{c}
\alpha\\
\beta
\end{array}
\right]
=
\left[
\begin{array}{c}
\theta_x\\
\theta_y
\end{array}
\right]
\;.
\eeq
Thus,

\beq
\left[
\begin{array}{c}
\alpha\\
\beta
\end{array}
\right]
=
\frac{1}{\Delta}
\left[
\begin{array}{cc}
s_{2y} & -s_{2x} \\
-s_{1y} & s_{1x}
\end{array}
\right]
\left[
\begin{array}{c}
\theta_x\\
\theta_y
\end{array}
\right]
\;.
\label{eq-ab-fun-theta}
\eeq
The determinant $\Delta$
is given by

\beq
\Delta = s_{1x} s_{2y} - s_{1y}s_{2x}=
\hats_1\times \hats_2 \cdot \hate_z = w_z
\;.
\eeq
Substituting the
expressions for $\alpha, \beta$
given by Eq.(\ref{eq-ab-fun-theta})
into the Z component of
Eq.(\ref{eq-def-vec-theta})
now yields

\begin{subequations}\beqa
\theta_z &=& \alpha s_{1z} + \beta s_{2z}\\
&=&
\left(
\frac{s_{2y} \theta_x - s_{2x}\theta_y}
{\Delta}\right)
s_{1z}
+
\left(
\frac{-s_{1y} \theta_x + s_{1x}\theta_y}
{\Delta}\right)
s_{2z}\\
&=&
-k_x \theta_x - k_y \theta_y
\;,
\eeqa\end{subequations}
where

\beq
k_\mu = \frac{
w_\mu}{
w_z}
\;
\eeq
for $\mu\in \{x, y\}$.

At this point, we have
reduced our problem to
the following 4 equations for the
4 unknowns
$\gamma, \theta_x, \theta_y, \theta_z$:

\begin{subequations}
\label{eq-gamma-theta-xyz-eqs}
\beq
|x|^2 = \cos^2 \theta
+
\left(
\frac{\theta_z}{\theta}
\right)^2
\sin^2\theta
\;,
\label{eq-gamma-theta-xyz-eq-a}
\eeq

\beq
\tan(\angle(x)-
\gamma)=
\frac{\theta_z \sin\theta}
{\theta \cos\theta}
\;,
\label{eq-gamma-theta-xyz-eq-b}
\eeq

\beq
\tan(\angle(y)-
\gamma )=
\frac{\theta_x }{\theta_y}
\;,
\label{eq-gamma-theta-xyz-eq-c}
\eeq
and

\beq
\theta_z =
-k_x \theta_x - k_y \theta_y
\;.
\label{eq-gamma-theta-xyz-eq-d}
\eeq
\end{subequations}

Define the following two shorthand symbols

\beq
t_x = \tan(\angle(x)-\gamma),
\;\;
t_y = \tan(\angle(y)-\gamma)
\;.
\eeq
Eqs.(\ref{eq-gamma-theta-xyz-eq-c})
and (\ref{eq-gamma-theta-xyz-eq-d})
yield

\beq
\left[
\begin{array}{cc}
1 & -t_y \\
-k_x & -k_y
\end{array}
\right]
\left[
\begin{array}{c}
\theta_x\\
\theta_y
\end{array}
\right]
=
\left[
\begin{array}{c}
0\\
\theta_z
\end{array}
\right]
\;.
\eeq
Thus,

\beq
\left[
\begin{array}{c}
\theta_x\\
\theta_y
\end{array}
\right]
=
\frac{-\theta_z}{k_y + k_x t_y}
\left[
\begin{array}{c}
t_y\\
1
\end{array}
\right]
\;.
\label{eq-theta-xy-in-z}
\eeq
Substituting the values
for
$\theta_x$ and $\theta_y$
given by Eq.(\ref{eq-theta-xy-in-z})
into the definition of
$\theta$  yields:

\beq
\frac{\theta_z}{\theta}=
\frac{k_y + k_x t_y}
{\sqrt{(k_y + k_x t_y)^2 + t_y^2 + 1}}
\;.
\label{eq-theta-z-in-kt}
\eeq
Eqs.(\ref{eq-gamma-theta-xyz-eq-a})
and (\ref{eq-gamma-theta-xyz-eq-b})
yield

\beq
\left[
\begin{array}{cc}
1 & \left(\frac{\theta_z}{\theta}\right)^2 \\
t_x^2 & -\left(\frac{\theta_z}{\theta}\right)^2
\end{array}
\right]
\left[
\begin{array}{c}
\cos^2\theta\\
\sin^2\theta
\end{array}
\right]
=
\left[
\begin{array}{c}
|x|^2\\
0
\end{array}
\right]
\;.
\eeq
Thus,

\beq
\left[
\begin{array}{c}
\cos^2\theta\\
\sin^2\theta
\end{array}
\right]
=
\frac{|x|^2}
{
\left(
\frac{\theta_z}{\theta}
\right)^2
(1 + t_x^2)
}
\left[
\begin{array}{c}
\left(
\frac{\theta_z}{\theta}
\right)^2\\
t_x^2
\end{array}
\right]
\;.
\eeq
Consider the two components of
the vector on the right hand side
of the last equation. They must
sum to one:

\beq
\frac{
[
\left(
\frac{\theta_z}{\theta}
\right)^2
+ t_x^2
]|x|^2
}{
\left(
\frac{\theta_z}{\theta}
\right)^2
(1+t_x^2)
}
=
1
\;.
\label{eq-pre-final-non-lin-gamma}
\eeq
Substituting the value for
$\frac{\theta_z}{\theta}$
given by Eq.(\ref{eq-theta-z-in-kt})
into Eq.(\ref{eq-pre-final-non-lin-gamma})
finally yields

\beq
(k_y + k_x t_y)^2
(1 + t_x^2) |y|^2=
(1+t_y^2) t_x^2 |x|^2
\;.
\label{eq-final-non-lin-gamma}
\eeq
As foretold,
in order to find $\gamma$ in terms of
$(x,y)$,
we must solve for the root $\gamma$
of a nonlinear equation,
Eq.(\ref{eq-final-non-lin-gamma}).

\section{Appendix: Optimum Axis \\
of $U(2)$-subset}\label{app-mini}

Let $\{U_b\}_{\forall b}$ be
a $U(2)$-subset.
Suppose that we express each
$U_b$ in the form

\beq
U_b = e^{i\eta_b}
e^{i\sigz \gamma_b}
e^{i(\sigsone \alpha_b
+
\sigstwo \beta_b)}(i\sigw)^{f(b)}
\;,
\eeq
where $\eta_b, \alpha_b, \beta_b, \gamma_b$
are real parameters, where the vectors
$\hats_1, \hats_2$, and
$\hatw=\hats_1\times\hats_2$ are orthonormal,
and where $f(b)$ is an indicator
function which maps the
set of all possible $b$ into $\{0,1\}$.
Of course, $e^{i\eta_b} = \sqrt{\det{U_b}}$.
Appendix
\ref{app-param} shows how to
find the parameters
$\alpha_b, \beta_b, \gamma_b$ for a given
$U_b e^{-i\eta_b}\in SU(2)$.
The goal of this appendix is to
solve the following minimization problem.
If the value of the parameters
$\alpha_b, \beta_b, \gamma_b$
and the vectors $\hats_1, \hats_2,\hatw$
are allowed to vary, while keeping
the vectors $\hats_1, \hats_2, \hatw$ orthonormal
and keeping all $U_b$ fixed,
find vectors $\hats_1, \hats_2, \hatw$
that are optimal, in the sense
that they minimize
a cost function.
The cost function
penalizes deviations of the diagonal
matrices $e^{i\gamma_b\sigz}$
away from the 2d identity matrix
$I_2$.
Any choice of
orthonormal vectors
$\hats_1, \hats_2$
will be called {\bf strong directions}
and $\hatw=\hats_1\times\hats_2$
will be called a {\bf weak direction},
or an
{\bf axis of the $U(2)$-subset}.
An axis that minimizes the cost function
will be called
the {\bf optimum axis of the $U(2)$-subset}.

In Appendix
\ref{app-param}, we
used a quantity $\vec{\theta}$
such that

\beq
e^{i(\sigsone \alpha_b+\sigstwo \beta_b)}=
e^{i\vec{\sigma}\cdot \vec{\theta}_b}
\;.
\eeq
Hence,

\beq
\vec{\theta}_b = \alpha_b \hats_1
+ \beta_b \hats_2
\;.
\label{eq-vec-thetab}
\eeq
In this appendix, we will
find it convenient to use additional
symbols $\vec{r}_b, p_q, q_b $,
$X_{b,1}$, and $X_{b,2}$
which satisfy

\beq
e^{i(\sigsone \alpha_b+\sigstwo \beta_b)}=
p_b + i\vec{\sigma}\cdot \vecr_b
\;,
\eeq

\beq
\vecr_b= (\hats_1 X_{b,1} +
\hats_2 X_{b,2})q_b
\;,
\eeq

\beq
p_b^2 + q_b^2 = 1
\;,
\eeq
and

\beq
X_{b,1}^2 + X_{b,2}^2 = 1
\;.
\eeq
Eq.(\ref{eq-vec-thetab})
expresses $\vec{\theta}_b$
in terms of the
``fundamental" variables
$(\alpha_b, \beta_b, \gamma_b, \hats_1, \hats_2)$.
Likewise, $\vec{r}_b, p_q, q_b $,
$X_{b,1}$, and $X_{b,2}$ can be expressed
in terms of these
fundamental variables
as follows:

\beq
\vecr_b =
\frac{
(\alpha_b \hats_1
+ \beta_b \hats_2)}{
\sqrt{\alpha_b^2 + \beta_b^2}}
\sin \sqrt{\alpha_b^2 + \beta_b^2}
\;,
\eeq

\beq
p_b = \cos
\sqrt{\alpha_b^2 + \beta_b^2}
\;,
\eeq

\beq
q_b = \sin
\sqrt{\alpha_b^2 + \beta_b^2}
\;,
\eeq

\beq
X_{b,1}=
\frac{\alpha_b}
{\sqrt{\alpha_b^2 + \beta_b^2}}
\;,
\eeq

\beq
X_{b,2}=
\frac{\beta_b}
{\sqrt{\alpha_b^2 + \beta_b^2}}
\;.
\eeq

For each $b$, define a
{\bf correction} ${\cal C}_b$ by

\beq
{\cal C}_b =
e^{i \sigz \gamma_b}
\;.
\eeq
We will use the simple matrix norm
$\|A\| = \tr(AA^\dagger)$
(i.e., the sum of the absolute value of
each $A$ entry).
We define the cost function
(Lagrangian) ${\cal L}$ for our
minimization problem to be
the sum over $b$ of
the distance between
${\cal C}_b$ and the
2d identity matrix $I_2$. Thus,

\begin{subequations}\beqa
{\cal L} &=&
\sum_b \| {\cal C}_b - I_2 \| \\
&=&\sum_b \tr[ 2 -
({\cal C}_b + {\cal C}_b^\dagger) ]\\
&=& 4 \sum_b (1 - \cos \gamma_b)
\;.
\eeqa\end{subequations}

The cost function
variation is

\beq
\delta {\cal L}=
4 \sum_b \sin(\gamma_b) \delta \gamma_b
\;.
\label{eq-dl-in-dgamma}
\eeq
The  variations
$\delta\gamma_b$ represent $N_\rvb$
degrees of freedom ({\bf dof's}),
but they are not independent dofs, as
they are subject to the following
constraints.
For all $b$, $U_b$
is kept fixed
during the variation of ${\cal L}$, so

\begin{subequations}
\label{eq-vari-contr}
\beq
\delta U_b =
(i \sigz \delta\gamma_b)U_b
+ e^{i\eta_b}e^{i\sigz\gamma_b}
(\delta p_b +
i \vec{\sigma}\cdot \delta\vecr_b)(i\sigw)^{f(b)}
+ U_b (i f(b)\vec{\sigma}\cdot\delta\hatw)=0
\;.
\label{eq-vari-contr-ub}
\eeq
(We've used the fact that
$f(b)\in \{0,1\}$).
The vectors $\hats_1$ and
$\hats_2$ are kept
orthonormal
(i.e., $\hats_j \cdot \hats_k = \delta(j,k)$
for all $j,k \in \{1,2\}$)
during the variation  of ${\cal L}$, so

\beq
\delta(\hats_j \cdot \hats_k)=0
\;
\label{eq-vari-contr-ortho}
\eeq
for $j,k \in \{1,2\}$.
Finally, the points
$(p_b, q_b)$ and $(X_{b1}, X_{b2})$
are constrained to lie on the unit circle, so

\beq
p_b\delta p_b + q_b \delta q_b = 0
\;,
\label{eq-vari-contr-pb}
\eeq
and

\beq
X_{b1}\delta X_{b1} + X_{b2}\delta X_{b2}=0
\;.
\label{eq-vari-contr-xb}
\eeq
\end{subequations}
Eq.(\ref{eq-vari-contr-ub})
represents $3N_\rvb$ constraints.
Eq.(\ref{eq-vari-contr-ortho})
represents 3 constraints.
Eq.(\ref{eq-vari-contr-pb})
and Eq.(\ref{eq-vari-contr-xb})
together represent  $2N_\rvb$ constraints.
Thus, Eqs.(\ref{eq-vari-contr})
altogether represent
$5N_\rvb + 3$ (scalar) equations
in terms the $5N_\rvb +6$
(scalar) unknowns
(the unknowns are: 3 components of $\delta \hats_1$,
3 components of $\delta \hats_2$,
and, for all $b$,
$\delta\gamma_b, \delta p_b, \delta q_b,
\delta X_{b1}, \delta X_{b2}$).
Therefore, there are really
only 3 independent dofs
within these $5N_\rvb +6$
variations. Next, we will
 express $\delta {\cal L}$
in terms of only 3 independent
variations
(for independent variations,
we will find
it convenient to use
$\hatw\cdot\delta \hats_1,
\hatw\cdot\delta \hats_2$ and
$\hats_1\cdot \delta \hats_2$).
Once $\delta {\cal L}$
is  expressed in this manner, we will
be able to set to zero
the coefficients of the
3 independent variations.

Eq.(\ref{eq-vari-contr-ub}) implies
the following 4 equations:
(we use the fact that
$p_b^2+\vecr_b^2 = 1$)

\begin{subequations}
\label{eq-du-expanded}
\beq
- r_{bz}\delta\gamma_b -
\frac{\vecr_b\cdot\delta\vecr_b}{p_b}
-f(b)\vecr_b\cdot\delta\hatw=0
\;,
\label{eq-du-expanded-singlet}
\eeq
and

\beq
\vec{h}_b\delta \gamma_b
+
\delta \vecr_b
+
\vec{\epsilon}_b
=0
\;,
\label{eq-du-expanded-trio}
\eeq
\end{subequations}
where

\beq
\vec{h}_b =
\left[
r_{by},
-r_{bx},
p_b
\right]^T
\;,
\eeq
and

\beq
\vec{\epsilon}_b=
f(b)(p_b\delta\hatw
-\vecr_b\times\delta\hatw)
\;.
\eeq
Eqs.(\ref{eq-du-expanded})
constitute 4 constraints,
but only 3 are independent. Indeed,
if one dot-multiplies
Eq.(\ref{eq-du-expanded-trio})
by $\vecr_b$,
one gets Eq.(\ref{eq-du-expanded-singlet}).
So let us treat
Eq.(\ref{eq-du-expanded-singlet}) as
a redundant statement and ignore it.
Dot-multiplying
Eq.(\ref{eq-du-expanded-trio})
by
$\hats_1, \hats_2$ and $\hate_z$
separately, yields the
following 3 constraints:

\begin{subequations}
\label{eq-3-constr-pre-drb-subst}
\beq
(\hats_k\cdot\vec{h}_b)\delta\gamma_b
+\hats_k\cdot\delta \vecr_b
+\hats_k\cdot\vec{\epsilon}_b=0
\;,
\eeq
for $j\in\{1,2\}$, and

\beq
p_b \delta\gamma_b + \delta r_{bz}
+
\epsilon_{bz} =0
\;.
\eeq
\end{subequations}

Now we proceed to express
$\delta \vecr_b$ in terms of
$\delta\hats_1$ and $\delta\hats_2$.
From the definition
$\vecr_b=
\sum_{j=1}^2 \hats_j X_{bj} q_b$,
we immediately obtain

\beq
\delta\vecr_b=
\sum_{j=1}^2 (\delta\hats_j) X_{bj} q_b +
\sum_{j=1}^2 \hats_j\delta (X_{bj} q_b)
\;.
\label{eq-drb}
\eeq
Hence

\begin{subequations}
\label{eq-drb-scalar-eqs}
\beqa
\hats_k\cdot \delta \vecr_b &=&
\sum_j (\hats_k\cdot
\delta \hats_j)
 X_{bj} q_b
+
\sum_j \hats_k\cdot\hats_j
\delta(X_{bj}q_b)\nonumber\\
&=&
\sum_j (\hats_k\cdot
\delta \hats_j)
 X_{bj} q_b
+
\delta(X_{bk}q_b)
\;,
\label{eq-drb-dot-sk}
\eeqa
and
\beq
\delta\vecr_{bz}=
\sum (\delta s_{jz}) X_{bj} q_b +
\sum s_{jz}\delta (X_{bj} q_b)
\;.
\label{eq-drb-z}
\eeq
\end{subequations}

If we substitute the
expressions for
$\hats_k\cdot \delta \vecr_b $ and
$\delta\vecr_{bz}$ given by
Eqs.(\ref{eq-drb-scalar-eqs}) into
Eqs.(\ref{eq-3-constr-pre-drb-subst}),
we get

\begin{subequations}
\label{eq-3-constr-aft-drb-sub}
\beq
(\hats_k\cdot \vec{h}_b)\delta \gamma_b
+
\sum_j(\hats_k\cdot\delta\hats_j)X_{bj}q_b
+\delta(X_{bk}q_b)
+\hats_k\cdot \vec{\epsilon}_b =0
\;,
\label{eq-3-constr-aft-drb-sub-1}
\eeq
and

\beq
p_b\delta\gamma_b
+ \sum_j (\delta s_{jz})X_{bj}q_b
+ \sum_j s_{jz}\delta(X_{bj}q_b)
+ \epsilon_{bz}=0
\;.
\label{eq-3-constr-aft-drb-sub-2}
\eeq
\end{subequations}
Substituting the expression for
$\delta(X_{bj}q_b)$ given by
Eq.(\ref{eq-3-constr-aft-drb-sub-1})
into Eq.(\ref{eq-3-constr-aft-drb-sub-2})
yields

\beq
A_b\;\delta \gamma_b = B_b
\;,
\eeq
where

\beq
A_b=
p_b -\sum_j s_{jz}\vec{h}_b\cdot\hats_j
\;,
\eeq
and

\beq
B_b=
-\sum_j(\delta s_{jz}) X_{bj}q_b
+ \sum_{j,k} s_{jz}
(\hats_j\cdot\delta\hats_k)
X_{bk}q_b
+ \sum_j s_{jz}\hats_j\cdot\vec{\epsilon}_b
\;.
\label{eq-bq}
\eeq
Thus,
\beq
\delta{\cal L}=
4\sum_b \sin(\gamma_b)\delta\gamma_b
= 4\sum_b \sin(\gamma_b)
\frac{B_b}{A_b}
\;.
\label{eq-del-L-in-B}
\eeq

We have succeeded in expressing
$\delta{\cal L}$ in term of the 9 variations
$\delta\hats_1, \delta\hats_2, \delta\hatw$ of
the strong and weak directions. But not all
of these 9 variations are independent
due to the orthonormality of
$\hats_1, \hats_2, \hatw$. Our
next goal is to express these
9 variations in terms of 3
that can be taken to be independent.

For $j\in \{1,2\}$,
$\hats_j\cdot\hatw=0$ so

\beq
\hats_j\cdot\delta\hatw= - \hatw \cdot\delta\hats_j
\;.
\eeq
Note that

\beq
\hats_1\cdot\hats_2\times \delta \hatw
=
(\hats_1\times\hats_2)\cdot\delta \hatw
= \hatw\cdot \delta \hatw =0
\;.
\eeq
Thus,

\beq
\hats_j\cdot\hats_k\times \delta \hatw=0
\;
\eeq
for any $j,k\in \{1,2\}$.
Hence,

\beq
\hats_j\cdot\vecr_b\times \delta \hatw=0
\;.
\eeq
It follows that

\beq
\hats_j\cdot\vec{\epsilon}_b=
-f(b)p_b\hatw\cdot\delta \hats_j
\;.
\label{eq-eps-dot-sj}
\eeq

Define $\lambda, \lambda_1, \lambda_2$ by
\beq
\lambda = \hats_1\cdot\delta\hats_2
\;
\eeq
and

\beq
\lambda_j=\hatw\cdot \delta\hats_j
\;
\eeq
for $j\in \{1,2\}$.
One can always expand
$\delta\hats_1$ and
$\delta\hats_2$ in the
orthonormal basis
$(\hats_1, \hats_2, \hatw)$.
The constraints $\delta(\hats_j\cdot\hats_k)=0$
for
$j, k\in \{1,2\}$,
force such expansions to be:

\begin{subequations}
\label{eq-dels-12-in-sw-basis}
\beq
\delta\hats_1 = -\lambda \hats_2 + \lambda_1\hatw
\;,
\eeq
and

\beq
\delta\hats_2 = \lambda \hats_1 + \lambda_2\hatw
\;.
\eeq
\end{subequations}

Using Eqs.(\ref{eq-eps-dot-sj}) and
(\ref{eq-dels-12-in-sw-basis}),
$B_b$ as given by
Eq.(\ref{eq-bq}) can be re-written as

\beq
B_b=-q_b w_z\sum_j
X_{bj}\lambda_j
-f(b)p_b\sum_j s_{jz}\lambda_j
\;.
\eeq
Substituting this expression for $B_b$
into Eq.(\ref{eq-del-L-in-B})
for $\delta{\cal L}$
gives a new expression for
$\delta{\cal L}$.
In the new expression for $\delta{\cal L}$,
we may set the coefficients of
$\lambda_1, \lambda_2, \lambda$
separately to zero.
This yields:

\beq
0=\sum_b
\frac{\sin\gamma_b[q_bw_zX_{bj}
+ f(b) p_bs_{jz}] }
{p_b - \sum_js_{jz}\vec{h_b}\cdot\hats_j}
\;,
\label{eq-fruit-of-optimiz}
\eeq
for $j\in \{1,2\}$.

Next, we want to
solve the 2 equations
Eqs.(\ref{eq-fruit-of-optimiz}) for
the direction $\hatw$.
As in Appendix \ref{app-param},
let $k_\mu = w_\mu/w_z$ for $\mu\in\{x,y\}$.
Then

\begin{subequations}
\label{eq-strong-in-kxy}
\beq
\vec{w}=
\frac{[k_x, k_y, 1]^T}{\sqrt{1+ k_x^2 + k_y^2}}
\;.
\eeq
We can always assume that
$s_{1z}=0$. If we do so, then

\beq
\hats_1=
\frac{[k_y, -k_x, 0]^T}{\sqrt{k_x^2 + k_y^2}}
\;,
\eeq
and

\beq
\hats_2 = \hatw\times \hats_1
\;.
\eeq
\end{subequations}
Suppose we denote the two constraints of
Eq.(\ref{eq-fruit-of-optimiz}) by
$F_1=0, F_2=0$.
These two constraints
depend on the set of variables $V=
\{\hats_1, \hats_2\}\cup
\{\alpha_b, \beta_b, \gamma_b\}_{\forall b}$.
Using Eqs.(\ref{eq-strong-in-kxy})
and the results of Appendix \ref{app-param},
the variables $V$ can all be expressed
in terms of $k_x, k_y$ and
$\{U_b\}_{\forall b}$. Thus what
we really have is
$F_j(k_x, k_y, \{U_b\}_{\forall b})=0$
for $j\in\{1,2\}$. These two equations
can be solved numerically for
the two unknowns $k_x, k_y$.

\end{document}